\documentclass[aps,twocolumn,superscriptaddress,floatfix,longbibliography]{revtex4-2}
\usepackage{amsmath,amssymb,amsthm}
\usepackage{physics}
\usepackage{amsfonts}
\usepackage{mathrsfs}
\usepackage{graphicx}
\usepackage{subfigure}
\usepackage{tabularx}
\usepackage{enumerate}
\usepackage{dcolumn}
\usepackage{bm}
\usepackage{xcolor}
\usepackage[normalem]{ulem}
\usepackage[colorlinks,linkcolor=blue,citecolor=blue,urlcolor=blue]{hyperref}

\usepackage{tikz}
\newcommand*\emptycirc[1][0.6ex]{\tikz\draw (0,0) circle (#1);} 
\newcommand*\fullcirc[1][0.6ex]{\tikz\fill (0,0) circle (#1);} 

\begin{document}

\title{Stark many-body localization  with  long-range  interactions}
\author{Xiang-Ping Jiang}\thanks{These authors contributed equally to this work.}
\affiliation{Zhejiang Lab, Hangzhou 311121, China}

\author{Rui Qi}\thanks{These authors contributed equally to this work.}
\affiliation{Beijing National Laboratory for Condensed Matter Physics, Institute of Physics, Chinese Academy of Sciences, Beijing 100190, China}
\affiliation{School of Physical Sciences, University of Chinese Academy of Sciences, Beijing, 100049, China}

\author{Sheng Yang}
\affiliation{Zhejiang Institute of Modern Physics and Department of Physics, Zhejiang University, Hangzhou 310027, China}

\author{Yayun Hu}
\email{hyy@zhejianglab.edu.cn}
\affiliation{Zhejiang Lab, Hangzhou 311121, China}

\author{Guangwen Yang}
\email{ygw@tsinghua.edu.cn}
\affiliation{Zhejiang Lab, Hangzhou 311121, China}
\affiliation{Department of Computer Science and Technology, Tsinghua University, Beijing, 100084, China}

\date{\today}

\begin{abstract}
In one-dimensional (1D) disorder-free interacting systems, a sufficiently strong linear potential can induce localization of the many-body eigenstates, a phenomenon dubbed as Stark many-body localization (MBL). In this paper, we investigate the fate of Stark MBL in 1D spinless fermions systems with long-range interactions, specifically focusing on the role of interaction strength. We obtain the Stark MBL phase diagrams by computing the mean gap ratio and many-body inverse participation ratio at half-filling. We show that, for short-range interactions, there is a qualitative symmetry between the limits of weak and strong interactions. However, this symmetry is absent in the case of long-range interactions, where the system is always Stark many-body localized at strong interactions, regardless of the linear potential strength. Furthermore, we study the dynamics of imbalance and entanglement with various initial states using time-dependent variational principle (TDVP) numerical methods. We reveal that the dynamical quantities display a strong dependence on the initial conditions, which suggests that the Hilbert-space fragmentation precludes thermalization. Our results demonstrate the robustness of Stark MBL even in the presence of long-range interactions and
offer an avenue to explore MBL in disorder-free systems with long-range interactions.
\end{abstract}

\maketitle

\section{Introduction}
Many-body localization (MBL), in which sufficiently strong disorder can localize an
interacting many-body system, is a generalization of Anderson localization~\cite{gornyi2005interacting,basko2006metal,pal2010many,altman2015universal,abanin2017recent,alet2018many}. MBL serves as a robust counterexample to thermalization, as it violates the predictions of the eigenstate thermalization hypothesis (ETH)~\cite{deutsch1991quantum,srednicki1994chaos,nandkishore2015many,abanin2019colloquium}. This results in the memory effect of the initial state in local observables and an area law entanglement in MBL phases. The signatures of disorder-induced MBL phases have been observed in various experimental platforms~\cite{schreiber2015observation,luschen2017observation,kohlert2019observation,sierant2022challenges} and can be characterized by essential quantities such as level statistics, entanglement entropy, and its real-time dynamical behaviours. Although the conventional MBL phenomenon is typically studied in disordered systems, extensive research has shown that certain features of MBL phases can also emerge in disorder-free systems subjected to a Stark linear potential. The phenomenon of localization in many-body eigenstates induced by a linear potential, often referred to as Stark MBL, is a widely studied topic~\cite{schulz2019stark,van2019bloch,zhang2021mobility,yao2021many,wei2022static,doggen2021stark}. Many recent experiments have been proposed to realize Stark MBL and demonstrate that a strong electric field can induce 
localization without disorder~\cite{taylor2020experimental,morong2021observation,wang2021stark,guo2021stark}.

In recent years, there has been an increasing interest in studying thermalization and MBL in systems with long-range (LR) hopping and LR interactions, in comparison to short-range (SR) coupling systems~\cite{yao2014many,barbiero2015out,burin2015many,gutman2016energy,wu2016understanding,nag2019many,tikhonov2018many,deng2018duality,roy2019self,deng2019one,maksymov2020many,sierant2019many,modak2020many,deng2020universal,kuwahara2021absence,yousefjani2023mobility,vu2022fermionic,huang2023incommensurate,huang2023statistics,yousefjani2023floquet,li2021hilbert,korbmacher2023lattice,khemani2020localization,yousefjani2023long,cheng2023many}. Many numerical investigations on MBL in one-dimensional (1D) systems with LR 
hopping have led to a consensus that MBL states cannot survive in such systems with power-law hopping $t\sim 1/r^{\alpha}$
when $\alpha <2$~\cite{burin2015many,gutman2016energy,nag2019many,wu2016understanding}. This suggests that the MBL states can be destroyed by sufficiently LR hopping. However, the studies on the effects of LR interactions $U\sim 1/r^{\alpha}$ on MBL phases have revealed interesting findings. Specifically, it has been observed that weak and intermediate LR interactions have a similar impact to SR interactions in MBL phases, while the localization behavior changes significantly for strong LR interactions~\cite{li2021hilbert,vu2022fermionic,korbmacher2023lattice,khemani2020localization}. On the other hand, the understanding of the role of interaction strength in Stark MBL systems with LR interactions remains unclear~\cite{bhakuni2020stability,chanda2022many,Plukin2022many}. This raises questions regarding the possibility of achieving a robust Stark MBL phase in the presence of LR interactions, as well as the impact of the interaction strength on the localization behavior of disorder-free systems.

Here we address these questions by studying a tunable LR interacting system with a linear potential, focusing on the role of interaction strength. We present a comprehensive analysis of the ergodic-Stark MBL transitions and obtain the associated interacting phase diagrams by computing the mean gap ratio and many-body inverse participation ratio (IPR). Our numerical results reveal a qualitative symmetry between the weak and strong SR interaction limits, whereas no such symmetry exists for LR interactions. The strong LR interactions induce additional constraints that cause the shattering of the Hilbert-space fragments, resulting in a Stark MBL transition that is independent of the linear potential strength.

In addition to quantitative analysis of spectral characteristics of the systems, we also utilize time-dependent variational principle (TDVP) numerical methods to simulate the dynamics of relevant physical quantities, namely density imbalance and entanglement entropy. Our findings clearly indicate discernible differences between the ergodic and Stark MBL regimes in lattice systems subjected to a linear potential. Notably, our results demonstrate a significant slowdown in the dynamics for both strong SR and LR interacting systems. These results suggest that Stark MBL is due to Hilbert-space fragmentation in disorder-free systems.

The organization of the paper is as follows. In Sec.~\ref{sec:model}, the model of the interacting Hamiltonian is presented. In Sec.~\ref{sec:phase}, we numerically calculate the mean gap ratio and many-body IPR to obtain the Stark MBL phase diagrams with SR and LR interactions. Then, we study the dynamics of the model and the results show a strong dependence on the initial 
states in Sec.~\ref{sec:dynamics}. Finally, we summarize our main findings and give a brief conclusion in Sec.~\ref{sec:conclusion}.

\section{The model Hamiltonian}\label{sec:model}
We consider a 1D system of spinless fermions with lattice size $ L $ and half-filling, which is described by the following Hamiltonian:
\begin{equation}\label{eq:Hamiltonian}
	H = t\sum_{i=1}^{L} (c_i^\dagger c_{i+1} + \text{H.c.}) +\sum_{i=1}^L V_i n_i +\frac{1}{2}\sum_{i,j} U_{i,j}n_in_j,
\end{equation}
where $t$ is the nearest-neighbor hopping, $ c_i^\dagger $ and $ n_i= c_i^\dagger c_{i}$ are the creation operator and the  number operator at site $ i $, respectively. In addition, $ V_i $ is the strength of the on-site potential and $ U_{i,j} $ is the two-body interaction between the site $ i $ and $ j $. Here we study the Stark MBL transition, thus the form of the on-site potential is written as
\begin{align}\label{eq:linearpotential}
	V_i = -\gamma i+\alpha\qty(\frac{i}{L})^{2},
\end{align}
with $ \gamma $ being the linear potential strength, and add $ \alpha $ to break the pure linearity, which is necessary for a direct comparison to the disorder-driven MBL\cite{schulz2019stark,van2019bloch}. We take $ \alpha=0.5 $ in our paper.
To achieve a clear contrast effect, we examine two distinct scenarios of electron-electron interaction: the short-range (SR) and the long-range (LR) interactions. They are given by\cite{vu2022fermionic,huang2023statistics}
\begin{equation}\label{eq:interaction}
	U_{i,j} =\left\{
	\begin{aligned}
		&U\delta_{i\pm 1, j}, &\text{SR case}\\
		&U\left(\dfrac{L}{\pi}\sin\dfrac{\pi\lvert i-j \rvert}{L}\right)^{-\kappa}, &\text{LR case}
	\end{aligned}
    \right.
\end{equation}
where $U$ is the strength of electron-electron interaction. In this paper, we impose  the system at half filling under the periodic boundary condition (PBC). For each type of interaction, we consider $\kappa = 1$ and present $2$D phase diagrams in the interaction-field ($U-\gamma$) plane using the exact diagonalization (ED) method.

\begin{figure*}[t]
\centering
    \hspace{0.0cm}
	\includegraphics[width=0.40\textwidth]{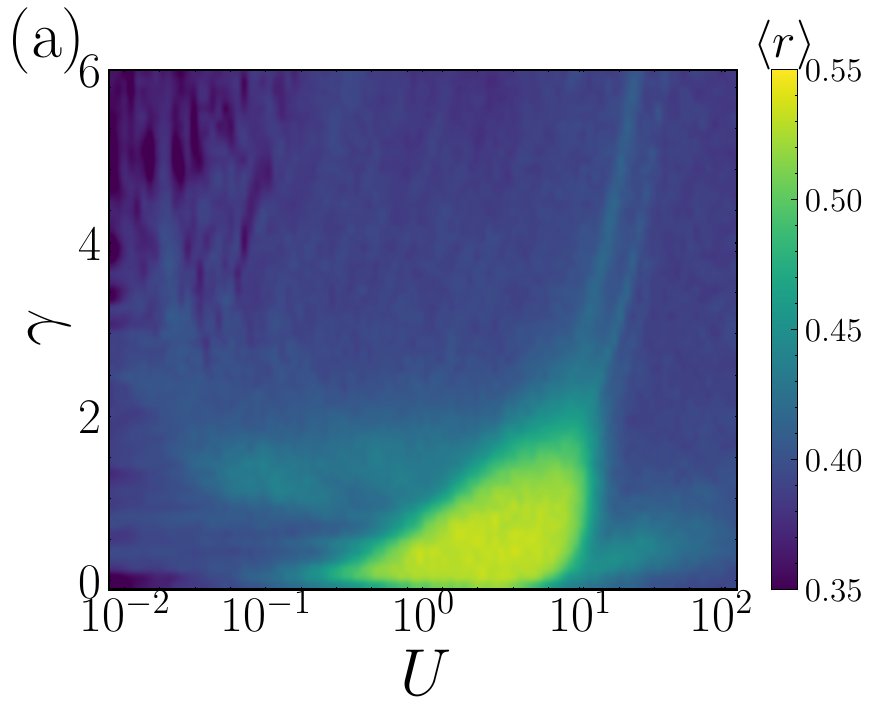}
	\vspace{-0.2cm}
	\includegraphics[width=0.40\textwidth]{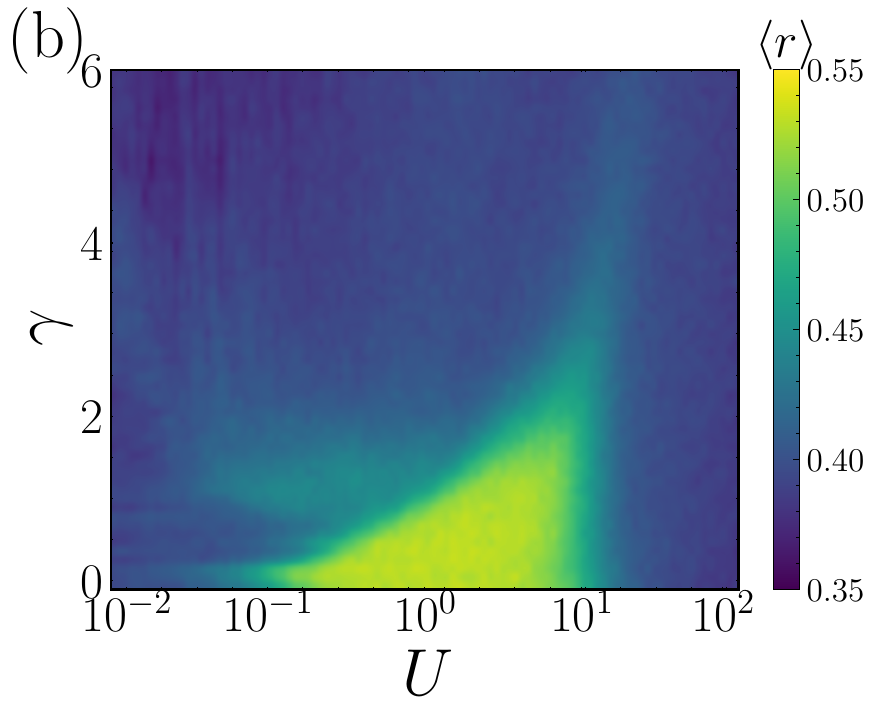}
	\includegraphics[width=0.40\textwidth]{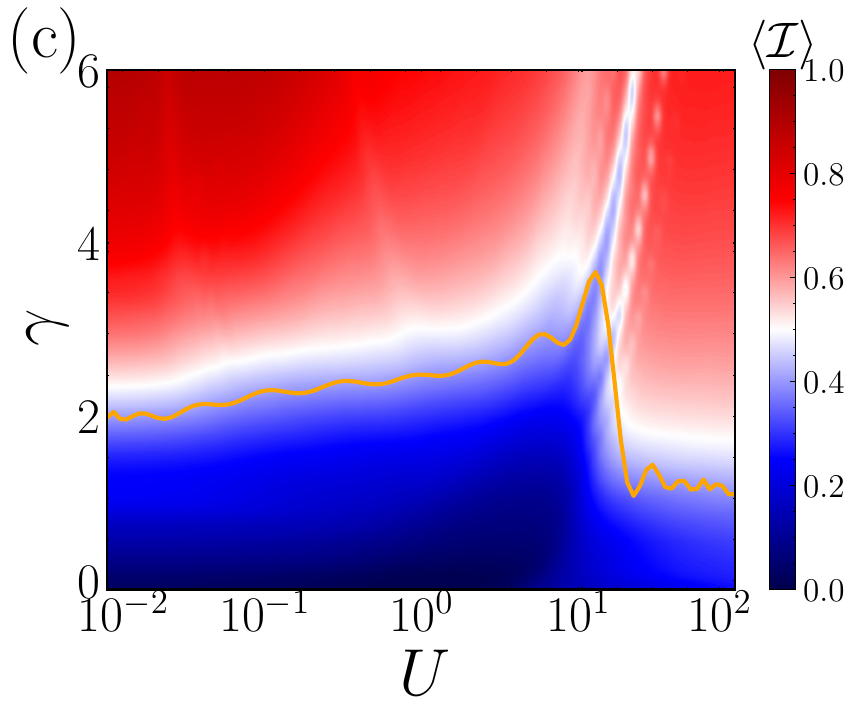}
	\hspace{0.0cm}
	\includegraphics[width=0.40\textwidth]{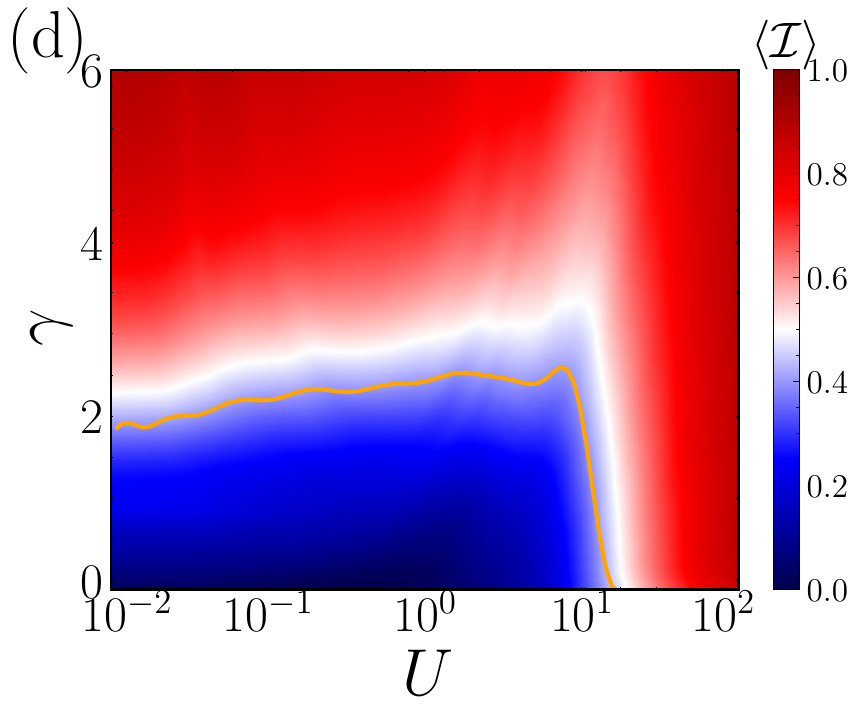}
	\vspace{-0.2cm}
	\caption{Left panel: The mean gap ratio $ \expval{r} $ and the MIPR $\expval{\mathcal{I}}$ of the SR interacting Hamiltonian. Right panel: The mean gap ratio $ \expval{r} $ and the MIPR $\expval{\mathcal{I}}$ of the LR interacting Hamiltonian with $\kappa=1$. The color represents the corresponding quantities. The orange lines mark the contour for $\expval{\mathcal{I}}=0.39$ in both SR and LR models. The value of $\expval{\mathcal{I}}$ is chosen because it corresponds to the localization transition point in each Hamiltonian (\ref{eq:Hamiltonian}) when $U = 0.01$. Here we consider a $L = 16$ system at half-filling under PBC.}
	\label{fig:phase}
\end{figure*}

\section{Interacting Phase diagram}\label{sec:phase}
In this section, we study the ergodic-Stark MBL transitions and obtain the Stark MBL phase diagrams with SR and LR interactions. Thus we consider two diagnostics: the mean gap ratio $ \expval{r} $ and the mean IPR $\expval{\mathcal{I}}$. We calculate the quantities based on ED of the Hamiltonian  (\ref{eq:Hamiltonian}) under PBC, and then obtain the Stark MBL interacting phase diagrams as shown in Fig. \ref{fig:phase}.
\subsection{Mean gap ratio}
One powerful and fundamental diagnostic tool for determining the localization properties of a system is spectral statistics. We calculate the mean gap ratio to distinguish the ergodic and Stark MBL phases. The gap ratio is defined by\cite{oganesyan2007localization,luitz2015many}
\begin{equation}\label{eq:gapratio}
	r_n = \min \left\{\frac{\delta E_{n}}{\delta E_{n+1}},\frac{\delta E_{n+1}}{\delta E_{n}} \right\},
\end{equation}
where $ E_{n} $ is the $n$-th eigenvalue and $ \delta E_{n}=E_{n+1}-E_{n} $ is the gap between two adjacent eigenenergies. The mean gap ratio $ \expval{r} $ is averaged over all eigenstates. It is shown that the MBL phase has an extensive set of quasilocal integrals of motion. 
In the MBL phase, the eigenstates with different eigenvalues of these integrals exhibit uncorrelated energy eigenvalues, resulting in a Poisson distribution and the mean gap ratio $\expval{r} \simeq 0.386$. However, in the ergodic phase, the strong repulsion between neighboring energy levels leads to statistical properties akin to those of the Gaussian orthogonal ensembles (GOE), characterized by a mean gap ratio $\expval{r} \simeq 0.530$. Figs. \ref{fig:phase}(a) and \ref{fig:phase}(b) display the mean gap ratio $ \expval{r} $ obtained from ED results for the half-filled Hamiltonian (\ref{eq:Hamiltonian}) with system size $L=16$. We exhibit Stark MBL phase diagrams of the Hamiltonian (\ref{eq:Hamiltonian}) with SR and LR interactions for a range of interaction strength $U$ in Figs. \ref{fig:phase}(a) and \ref{fig:phase}(b), respectively. 
We find that the phase diagram characterized by the mean gap ratio can be divided into three distinct regimes: the single-particle regime (weak interaction $U<U_{c1}$), the many-body  regime  (intermediate interaction $U_{c1}<U<U_{c2}$), and the Mott  regime  (strong interaction $U>U_{c2}$). We estimate that $U_{c1}\sim 0.1$ and $U_{c2}\sim 10$ from the phase diagrams [Fig. \ref{fig:phase}(a) and \ref{fig:phase}(b)].

In the weak interaction $U<U_{c1}$ regime, the SR and LR interactions are too weak to thermalize the many-body eigenstates. We always have $ \expval{r}\simeq 0.386 $ for all $\gamma$ and weak interactions. Therefore, this regime is characterized by the predominance of single-particle properties. In the intermediate interaction $U_{c1}<U<U_{c2}$ regime, on the other hand, the system exhibits a different spectral statistics behaviour compared to the single-particle regime. The ergodic phase appears in this many-body regime as indicated by $ \expval{r}\simeq 0.530$. In the strong interaction $U>U_{c2}$ regime, it
seems that $ \expval{r}$ for both systems don't exhibit a clear Stark MBL transition.
This is due to the fact that the Hilbert-space in this Mott regime is highly fragmented, leading to the mean gap ratio $\expval{r}$ combining the spectra of various invariant subspaces. As a result, the spectrum seems to be uncorrelated. The numerical phase diagrams obtained from our disorder-free Hamiltonian  (\ref{eq:Hamiltonian}) are similar to the observations made in disorder-driven MBL systems~\cite{li2021hilbert,vu2022fermionic,huang2023incommensurate,huang2023statistics}.

\subsection{Inverse participation ratio}
In addition to energy spectrum statistics, another quantity that can describe the localization property of a system is many-body IPR. We exactly diagonalize finite-size Hamiltonian (\ref{eq:Hamiltonian}) at the half-filling, then each eigenstate $\psi_{\lambda}$ ($\lambda$ is the label of many-body eigenstates) can be quantified by the many-body IPR, which is defined as~\cite{bera2015many,vu2022fermionic,huang2023incommensurate,huang2023statistics}
\begin{equation}\label{eq:IPR}
	\mathcal{I}^{(\lambda)} = \dfrac{1}{1-\nu}\qty(\frac{1}{N_e}\sum_{i=1}^L \lvert{u_i^{(\lambda)}}\rvert^2-\nu), 
\end{equation}
where $u_i^{(\lambda)} \equiv \expval{n_i}{\psi_{\lambda}}$ is the projected particle number on site $i$ and the filling factor $\nu=N_e/L$ fixed at $1/2$. 
One can see that $\mathcal{I}\rightarrow 0$ for an extended state, while $\mathcal{I}\rightarrow 1$ for a localized state. Therefore, we have $0\le \mathcal{I} \le 1$ for a generic eigenstate. It should be noted that our study covers both weak and strong interactions, allowing us to examine the entire spectrum and employ the arithmetic mean IPR (MIPR), $\expval{\mathcal{I}} \equiv \sum_\lambda \mathcal{I}^{(\lambda)}/\mathcal{D}$, where $ \mathcal{D}=\binom{L}{N_e}$ is the dimension of the Hilbert space. Hence, the MIPR  $\expval{\mathcal{I}} $ (averaged over all eigenstates) characterizes the localization properties of a system. For simplicity, we will refer to $\expval{\mathcal{I}} $ simply as the MIPR.

By the exact diagonalization of the Hamiltonian (\ref{eq:Hamiltonian}) with SR and LR interactions at half filling, we compute the MIPR $\expval{\mathcal{I}} $ as a function of $\gamma$ and $U$ for system size $L=16$ to obtain the phase diagram. The calculated Stark MBL phase diagrams are presented in Figs. \ref{fig:phase}(c) and \ref{fig:phase}(d).
It is well established that the noninteracting Wannier-Stark model exhibits a localization transition point at $\gamma=2$, where all eigenstates are localized (extended) for $\gamma>2$ ($\gamma<2$) \cite{wannier1962dynamics,emin1987existence,qi2023localization}.
As shown in Figs. \ref{fig:phase}(c) and \ref{fig:phase}(d), this noninteracting localization persists for both SR and LR interacting systems in the regime of extremely weak interactions ($U\ll U_{c1}$), aligning with the assertion that MBL exists under perturbatively weak interaction conditions~\cite{basko2006metal,nag2019many}.

When $U\sim \mathcal{O}(1)$, we turn to the intermediate interaction regime. In this regime, there is a transition from the ergodic phase to the Stark MBL phase, which is relatively insensitive to the type of interaction. Specifically, in the extrapolated thermodynamic limit, the ergodic phase extends to the critical field strength $\gamma_{c} \sim 2.5$ for both the SR and LR interaction models, above which the corresponding phase is always Stark MBL. Though similar phase diagrams can be
obtained in disorder-induced MBL systems with LR interactions~\cite{vu2022fermionic,huang2023statistics}. It is important to highlight essential differences between Stark MBL and disorder-driven MBL. In the former case, localization arises from the Hilbert-space fragmentation ~\cite{li2021hilbert,doggen2021stark}, whereas in the latter case, it is a consequence of the emergence of local integrals of motion. 

\begin{figure}[b]
\centering
	\includegraphics[width=1.0\columnwidth]{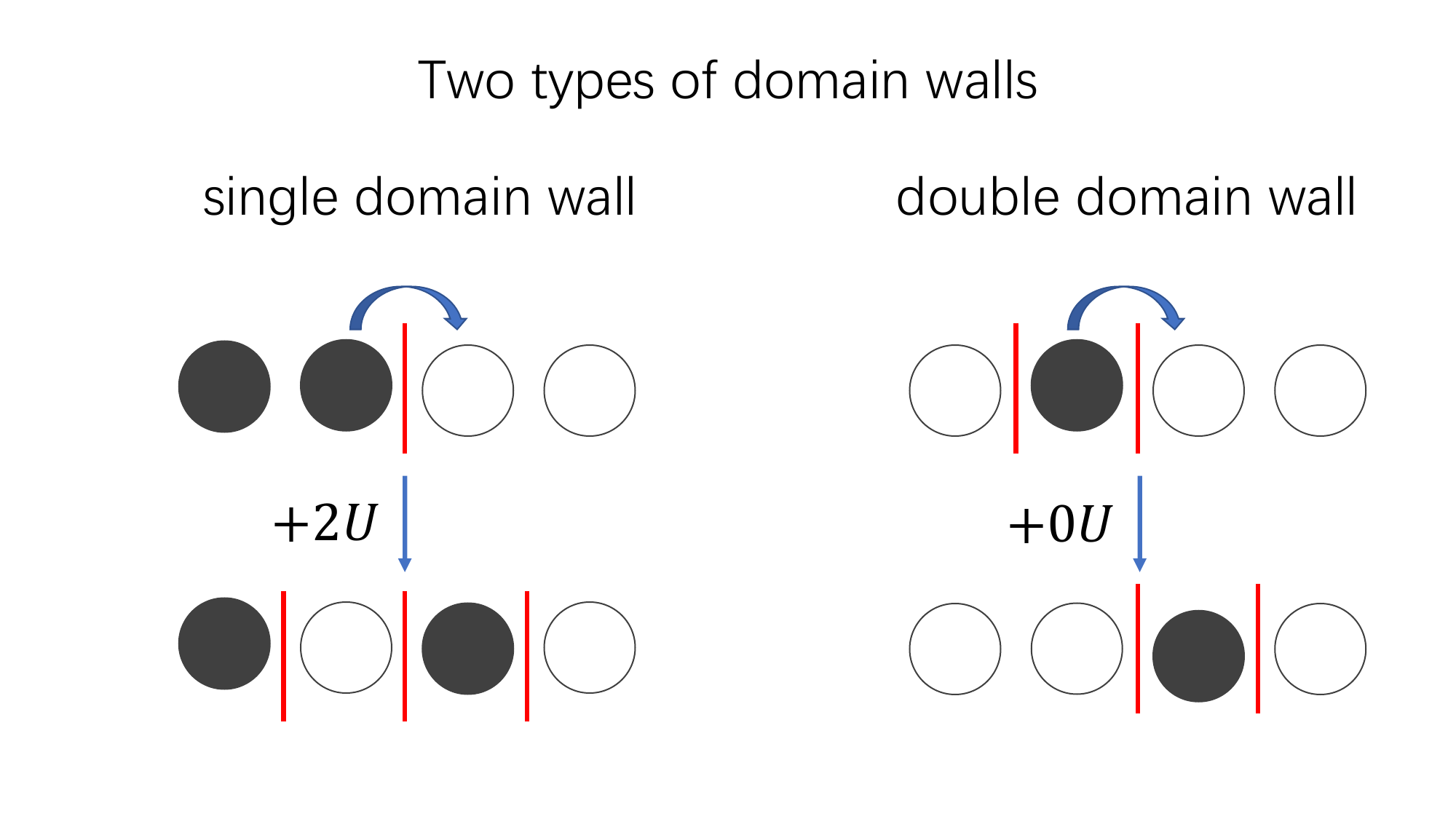}
	\caption{Schematic diagram of two domain walls. The left and right planes are single and double domain wall, respectively. }
	\label{fig:2}
\end{figure}

Up to the intermediate region, we observe no fundamental differences between the SR and LR interacting systems. However, this is no longer true in the strongly interacting regime ($U>U_{c2}$). This suggests that, in the context of LR interactions, the characteristics of the strongly interacting regime are distinct from those of the noninteracting regime. Indeed, the transition from the ergodic to the interaction-driven MBL at $U_{c2}$ exhibits a sharp phase transition as shown in Fig.~\ref{fig:phase}(d). The localization of many-body eigenstates is solely driven by interactions when $U \gg U_{c2}$, reminiscent of the mechanism observed in a Wigner crystal. In this Mott regime, the LR interactions are found to result in Hilbert space fragmentation and localization, independent of the linear potential strength. The similar phenomenon is also found in LR interacting systems with disorder\cite{vu2022fermionic,huang2023statistics,khemani2020localization,li2021hilbert,korbmacher2023lattice}. In the limit of strong-$U$, only Fock basis states with the same interaction energies can hybridize. However, for LR interactions, these states can only be connected through a global translation operator, meaning that the states are related by shifting all occupied sites by the same amount. This phenomenon indicates that the localization is predominantly influenced by the interaction between particles. We would like to emphasize that this interaction-driven MBL phase is exclusively present in the LR interaction case for the strong-$U$ limit, thereby conclusively establishing it as a strongly correlated phenomenon.  

However, compared with LR interactions, this picture is completely different for SR interactions, as there apparently exists an ergodic to Stark MBL phase transition in the $U>U_{c2}$ Mott regime [see Fig.~\ref{fig:phase}(c)]. Obviously, the linear potential $\gamma$ must play a critical role in the phase transition for SR interactions. In the regime of strong interactions, the presence of SR interactions does not completely eliminate the direct coupling terms between Fock basis states. We can use a argument picture to explain the phenomenon. As shown in Fig.~\ref{fig:2}, we consider two types of domain walls: a single domain wall (SDW) with no other nearby domain walls, and a double domain wall(DDW) consisting of two domain walls adjacent to each other. When a single electron hops to a SDW, it can create two additional domain walls in the process. This violates the conservation of domain wall number and an SDW is immobile when $U \rightarrow \infty$. On the other hand, the movement of the DDW preserves the overall domain wall configuration, maintaining the same number of domain walls in the system. A DDW can propagate without interference throughout the system, similar to a free particle in a noninteracting system. Thus, the above simple picture suggests a qualitative symmetry between weak-$U$ and strong-$U$ limits for strongly SR interacting systems.

\begin{figure}[h!]
\centering
	\includegraphics[width=0.9\columnwidth]{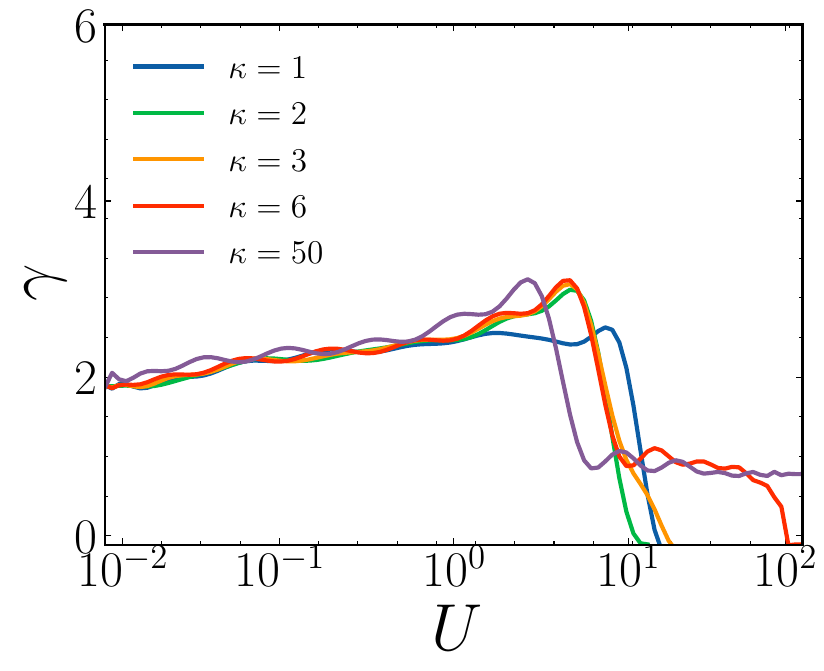}
	\caption{Schematic diagram of phase boundaries with different decay exponent $\kappa$ for a fixed value of $\expval{\mathcal{I}}=0.35$, such that the boundaries originate at $\gamma=2$ when $U=0$. Here we consider $L = 16$ systems at half-filling under PBC.}
	\label{fig:kappa}
\end{figure}

In the strongly interacting regime, we presented the DW argument for SR interaction and the Fock state splitting for LR interaction. These results can be understood as forms of Hilbert-space fragmentation\cite{yang2020hilbert,sala2020ergodicity,de2019dynamics,korbmacher2023lattice,khemani2020localization,li2021hilbert}. The notable distinctions between the LR and SR cases arise from the fact that the coupling within each fragment is suppressed by $U$ in the LR case, whereas in the SR case it is the original electron hopping. Therefore, it is natural to ask whether our findings in the strong interaction limit can be applied to other types of interactions. In order to answer this question, we now examine a more general form of interaction with $1 \leq \kappa < \infty $ in Eq.~(\ref{eq:interaction}). The numerical results are presented in Fig.~\ref{fig:kappa}. We find that the phase boundaries with different $ \kappa $ for $L=16 $ coincide with the SR case ($ \kappa \gg 1 $) until a critical value $U_{c2}^{\kappa}$ then eventually reduce to zero, similar to the LR case ($ \kappa = 1 $). Hence, the two cases examined in this study can be considered as two limiting scenarios, with various other cases of generic interactions existing between them.

\begin{figure}[b]
\centering
	\includegraphics[width=1.0\columnwidth]{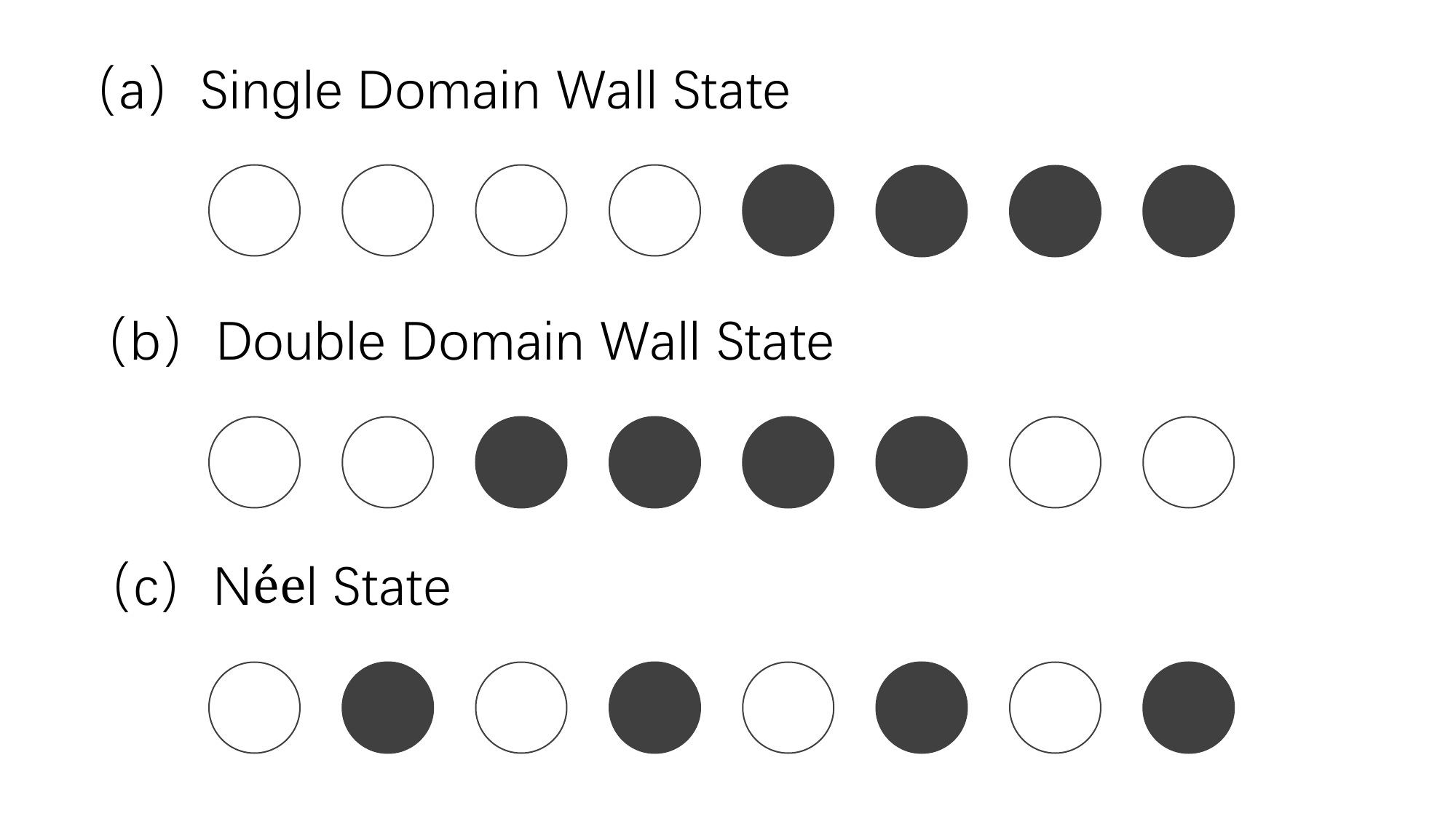}
	\caption{Schematic diagram of three different initial states for the system with lattice length $L = 8$. (a)-(c) denote the single domain wall initial state,  the double domain wall initial state and the N{\'e}el initial state, respectively.}
	\label{fig:3}
\end{figure}

\begin{figure*}[t]
	\centering
	\includegraphics[width=0.8\textwidth]{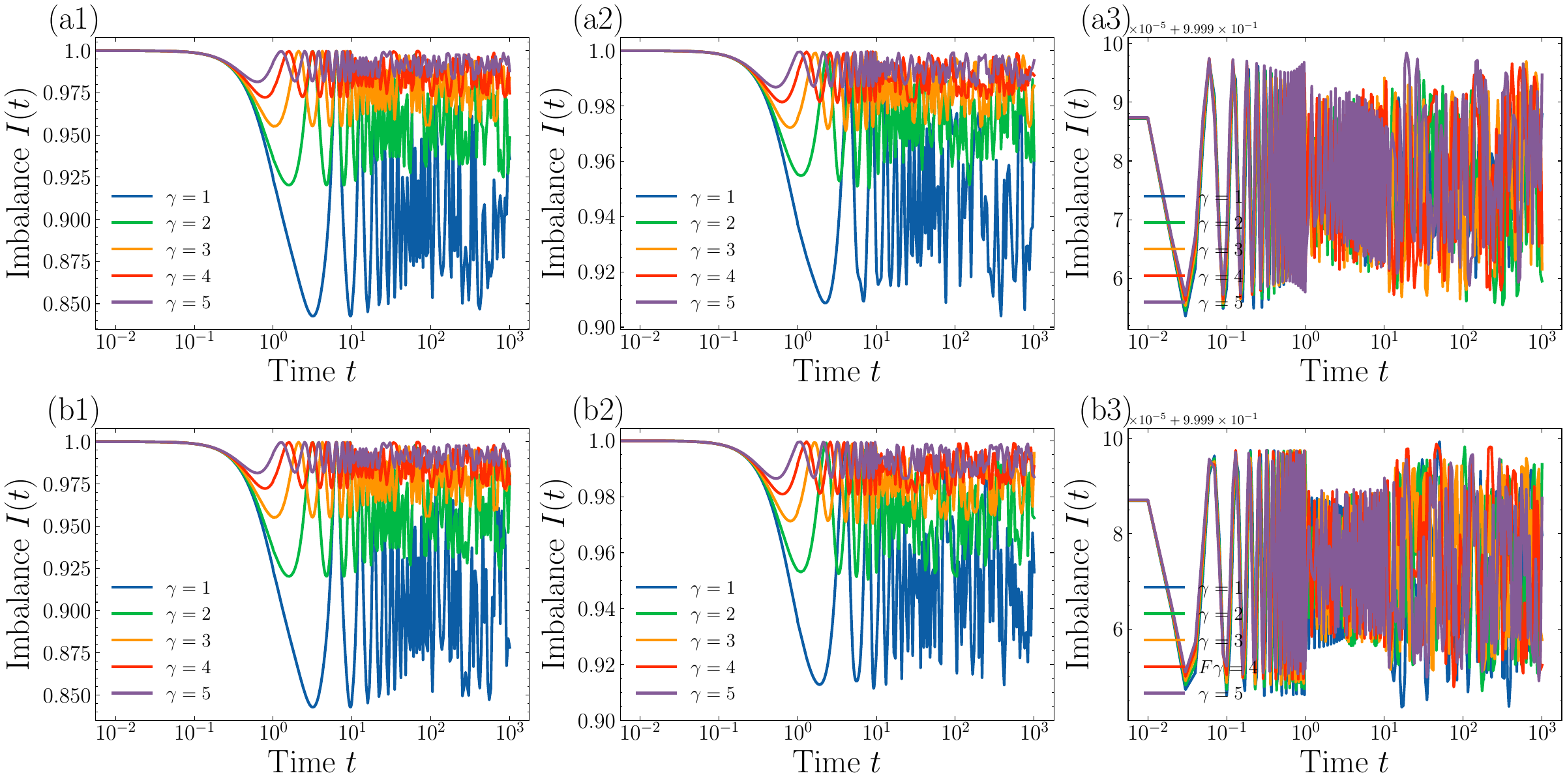}
	\includegraphics[width=0.8\textwidth]{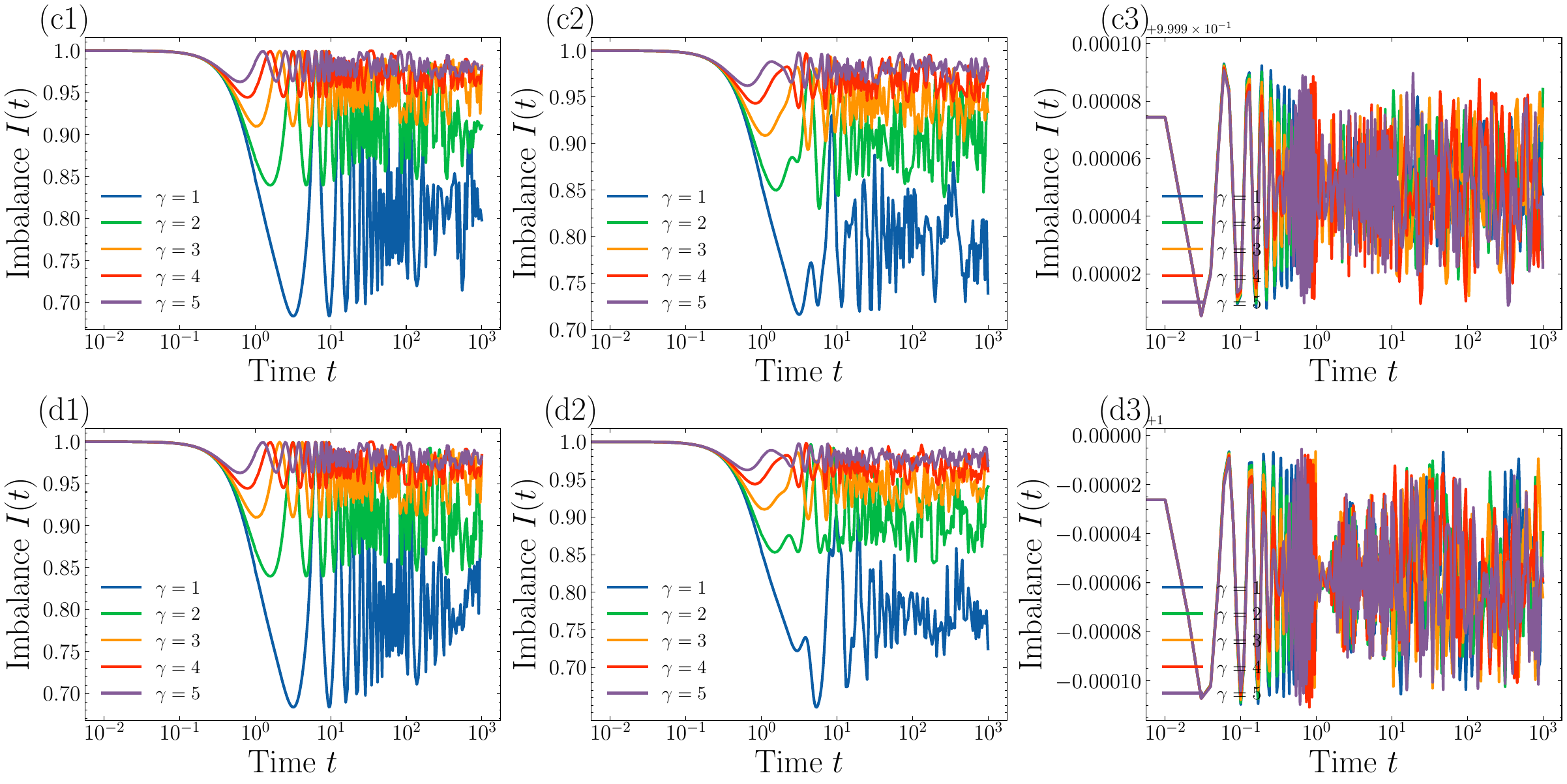}
	\caption{The imbalance dynamics $ I(t) $ as a function of time $t$ for various linear potential strengths $\gamma=1,2,3,4,5$. (a1)-(a3) and (b1)-(b3) show the results of SR and LR interacting systems for SDW initial sate, respectively. (c1)-(c3) and (d1)-(d3) show the results of SR and LR interacting systems with DDW initial sate, respectively. From left to right, the strength of interacting is $U = 0.01, 1$, and $100$. We consider a system of size $L = 48$ under OBC. All numerical data have been smoothed by the convolution.}
	\label{fig:imbalance}
\end{figure*}

\begin{figure*}[t]
	\centering
	\includegraphics[width=0.8\textwidth]{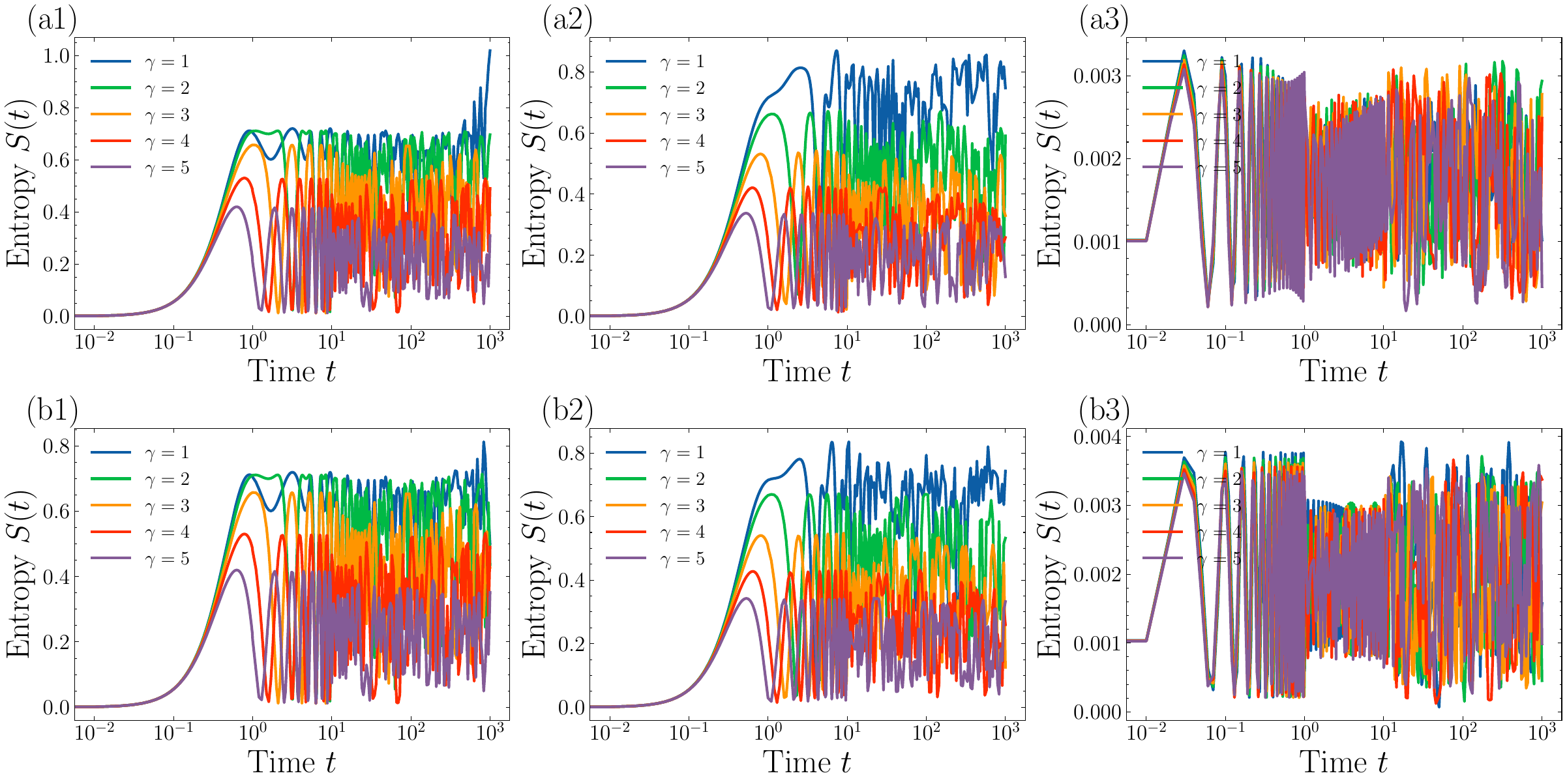}
	\includegraphics[width=0.8\textwidth]{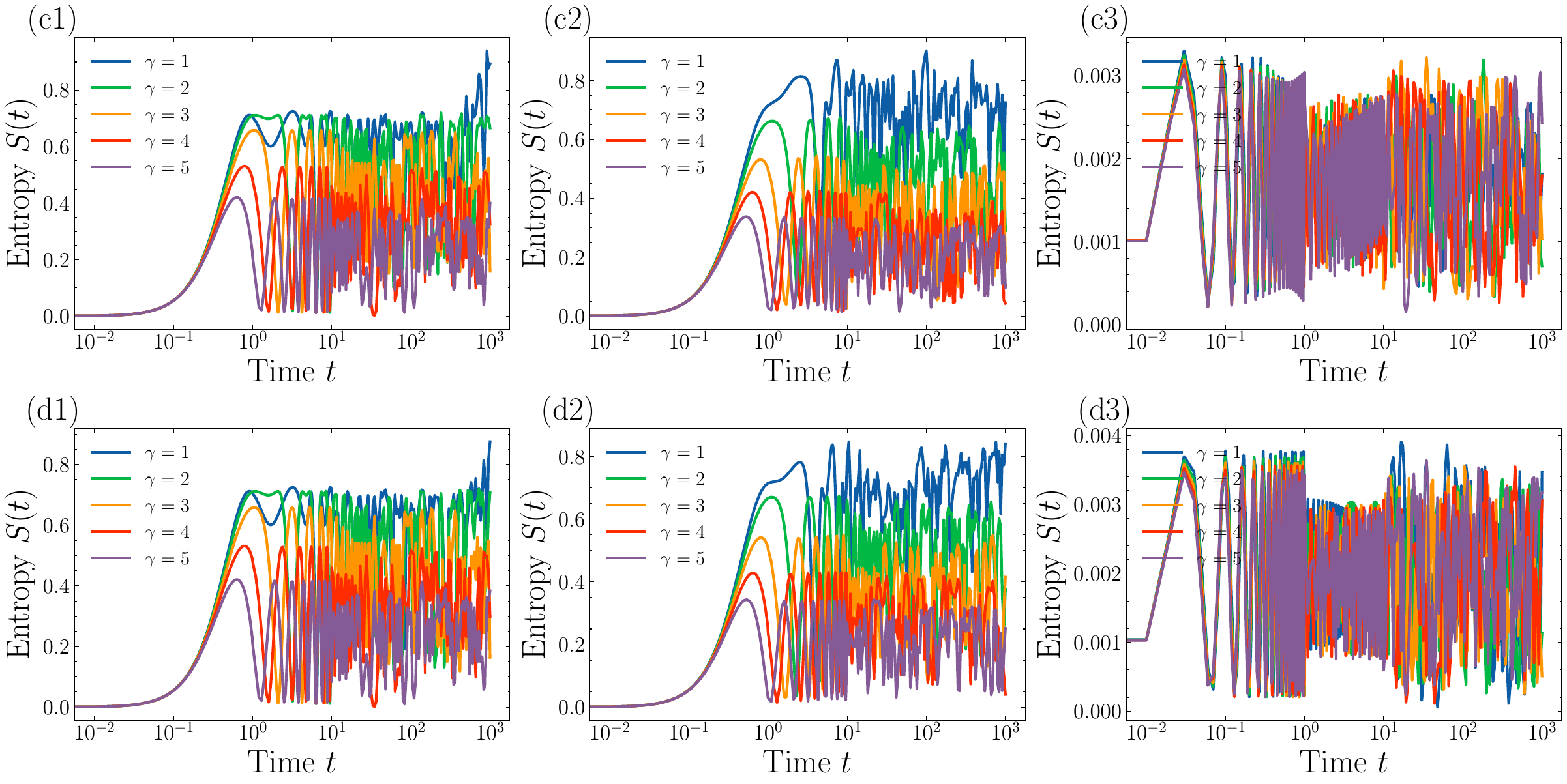}
	\caption{The entanglement entropy dynamics $ S(t) $ as a function of time $t$ for various linear potential strengths $\gamma=1,2,3,4,5$. (a1)-(a3) and (b1)-(b3) show the results of SR and LR interacting systems for SDW initial sate, respectively. (c1)-(c3) and (d1)-(d3) show the results of SR and LR interacting systems with DDW initial sate, respectively. From left to right, the strength of interacting is $U = 0.01, 1$, and $100$. We consider a system of size $L = 48$ under OBC. All numerical data have been smoothed by the convolution.}
	\label{fig:entropy}
\end{figure*}

\begin{figure*}[t]
	\centering
	\includegraphics[width=0.8\textwidth]{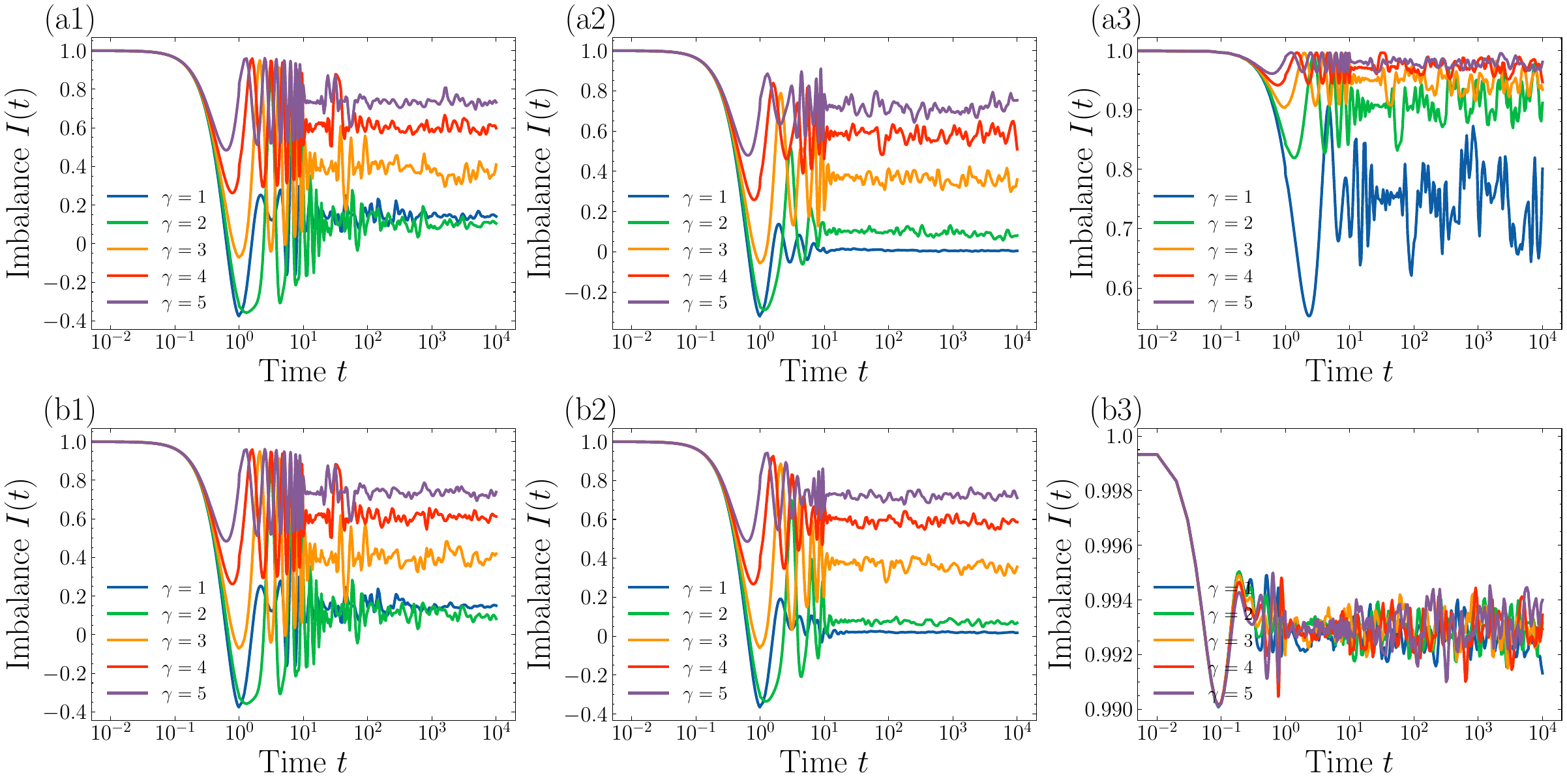}
	\includegraphics[width=0.8\textwidth]{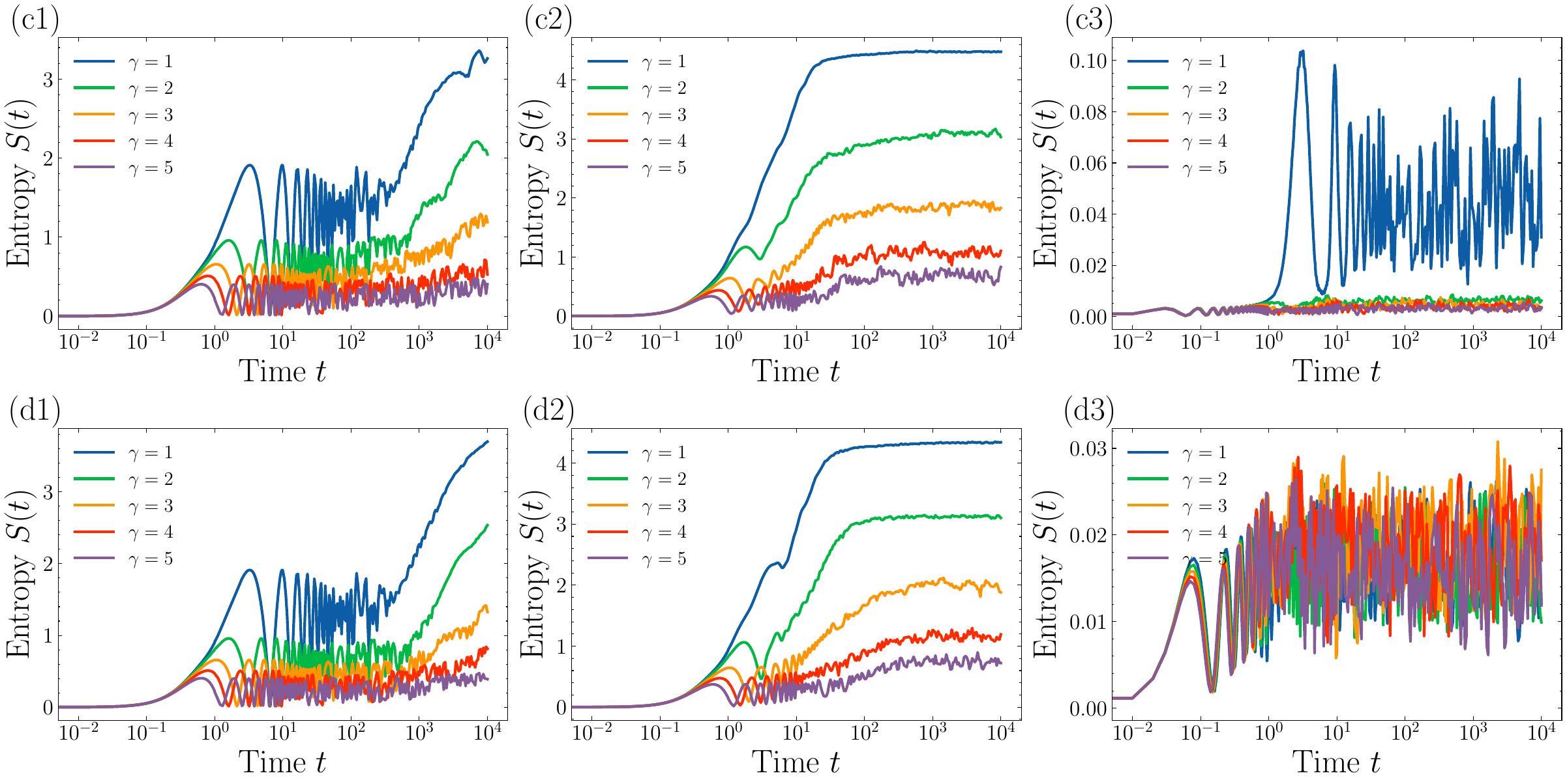}
	\caption{The dynamics of imbalance $ I(t) $ and entanglement entropy  $ S(t) $ as a function of time $t$ for N{\'e}el initial sate. (a1)-(a3) and (b1)-(b3) show the imbalance dynamics results of SR and LR interacting systems, respectively. (c1)-(c3) and (d1)-(d3) show the entanglement dynamics results of SR and LR interacting systems, respectively. From left to right, the strength of interacting is $U = 0.01, 1$, and $100$. We consider a system of size $L = 18$ under OBC. All numerical data have been smoothed by the convolution.}
	\label{fig:imbalance_entropy}
\end{figure*}

\section{Imbalance and entanglement dynamics}\label{sec:dynamics}
To study the dynamics of imbalance and entanglement in the model (\ref{eq:Hamiltonian}) with tunable strengths for SR and LR interactions, we employ a powerful numerical approach to characterize the temporal dynamics in 1D lattice systems. This method is based on the time-dependent variational principle (TDVP) as applied to matrix product states (MPS)~\cite{haegeman2011time,haegeman2016unifying}. Due to the presence of LR interactions in the Hamiltonians of our models, the TDVP becomes particularly advantageous for simulating dynamics. We perform the corresponding tensor-network calculations by means of the TDVP numerical method in the present work. The performance of the TDVP has been extensively employed to investigate the dynamics in MBL systems~\cite{Plukin2022many,haegeman2011time,haegeman2016unifying,doggen2021many}. For a comprehensive review of MPS-based methods for simulating dynamics, we refer the reader to the Ref.~\cite{paeckel2019time}. Implicitly, time evolution due to
the TDVP is described by
\begin{equation}
    \frac{d}{dt}\ket{\psi}=-i{\mathcal{P}}_{\rm{MPS}}H\ket{\psi},
\end{equation}
where $ \ket{\psi} $  is the time-dependent wave function and $ {\mathcal{P}}_{\rm{MPS}} $ projects the time-evolved wave function back onto the variational MPS manifold, with a dimension significantly smaller than the dimension of the full Hilbert space $\mathcal{D}$.

We consider the Hamiltonian (\ref{eq:Hamiltonian}) under open boundary conditions (OBC) and compute two dynamical quantities: density imbalance and entanglement entropy. As shown in Fig.~\ref{fig:3}, we consider three different initial states: SDW state, DDW state, and N{\'e}el sate. In this paper, We use a numerical time step of $\delta t = 0.1$ and a bond dimension $\chi = 512 $, which determines the size of the variational manifold. For SDW and DDW initial states, we compute dynamics up to $t = 10^{3}$ with the system size $L=48$. For N{\'e}el initial state, we perform the calculations up to $t = 10^{4}$ with the system size $L=18$. 
In the following, we show that the dynamics of imbalance and entanglement exhibit a significant dependence on the initial conditions. This observation suggests that Hilbert-space fragmentation prevents thermalization. Thus, the localization in disorder-free systems differs from disorder-induced MBL, in agreement with recent predictions based on localization via Hilbert-space shattering~\cite{li2021hilbert,doggen2021stark}.

\subsection{Density imbalance dynamics}\label{sec:imbalance dynamics}
We now provide dynamical evidence that the SR and LR interacting models are qualitatively similar in the weak interaction single-particle and intermediate interaction many-body regime. We initially examine the imbalance dynamics across the ergodic-Stark MBL transition. The imbalance measures the degree of memory retention of the initial state at the time $t$. It is defined as
\begin{equation}\label{eq:imbalance}
	{I}(t) = \dfrac{1}{N}\sum_{i=1}^L(-1)^{i}\expval{n_i}{\Psi_{0}(t)}, 
\end{equation}
From Eq.~(\ref{eq:imbalance}), it is easy to know that the initial value of the imbalance ${I}(t=0) = 1$. We numerically simulate the dynamics of three initial states as shown in Fig.~\ref{fig:3}, namely  SDW initial state $ \ket{\Psi_{0}}  = |\cdots \emptycirc\,\emptycirc\,\emptycirc\,\emptycirc\, \fullcirc\,\fullcirc\,\fullcirc\,\fullcirc\ \cdots \rangle $, DDW initial state $|\cdots \emptycirc\,\emptycirc\,\fullcirc\,\fullcirc \cdots \fullcirc\,\fullcirc\,\emptycirc\,\emptycirc \cdots \rangle $, and N{\'e}el initial state $ |\cdots \emptycirc\,\fullcirc\,\emptycirc\,\fullcirc\, \emptycirc\,\fullcirc\,\emptycirc\,\fullcirc \cdots \rangle $ (where $\emptycirc$ corresponds to an empty dot, and $\fullcirc$ to an occupied dot). For the above-specified initialization of the wave function, the state evolve with respect to the Hamiltonian (\ref{eq:Hamiltonian}) can be obtained as $ \ket{\Psi_{0}(t)}=e^{-iHt}\ket{\Psi_{0}} $.

During the course of time evolution, it is commonly observed that imbalances tend to decrease and converge towards a constant value. Subsequently, they tend to oscillate around this value, exhibiting relatively small amplitudes. In the ergodic phase, the constant value is typically close to zero. Conversely, in the MBL phase, this value remains relatively large, suggesting the presence of some memory from the initial state. Therefore, the long-time asymptotic behavior of imbalance can be used as a reliable indicator of localization~\cite{prasad2022initial,gunawardana2022dynamical,zisling2022transport}. 

Several representative cases of the imbalance dynamics ${I}(t)$ are depicted in Fig.~\ref{fig:imbalance}. We show the dynamics of imbalance
for both SR and LR interacting cases using the parameters  $\gamma=1,2,3,4,5$ and 
$U=0.01,1,100$. Figs.~\ref{fig:imbalance}(a1-a3) and (b1-b3) depict the results for the SDW initial condition, while Figs.~\ref{fig:imbalance}(c1-c3) and (d1-d3) correspond to the DDW initial condition. In Figs.~\ref{fig:imbalance}(a1-a2) and (c1-c2), the ${I}(t)$ decrease to a large stable value, even for a weak field gradient $\gamma=1$. 
Our findings reveal that initial states with different energy (the SDW and DDW configurations) can exhibit locally identical dynamics over extended time periods. Especially, in the strong interaction Mott regime, the ${I}(t)$ almost not decay and remains at its initial value. Therefore, the imbalance dynamics for the initial states of SDW and DDW are practically indistinguishable across the entire parameter region examined in this study. 

However, as shown in Figs.~\ref{fig:imbalance_entropy} (a1-a3) and (b1-b3), we find that the imbalance dynamics ${I}(t)$ for the N{\'e}el initial condition is very different from the results in Fig.~\ref{fig:imbalance}. In Figs.~\ref{fig:imbalance_entropy} (a1-a2) and (b1-b2),  the imbalance decays exponentially fast to zero in the ergodic phase for linear potential $\gamma \leq \gamma_{c}$, while in the Stark MBL phase the imbalance tends to a finite value. Meanwhile, in Figs.~\ref{fig:imbalance_entropy} (a3) and (b3), we see that the asymptotic value of SR interacting system is smaller than LR interacting system, which the value is very close to $1$. This observation suggests that a system with strong LR interactions undergoes Hilbert-space fragmentation, thereby inhibiting strong thermalization. In the context of strong ETH, this implies that the long-time behavior of local observables remains invariant regardless of the initial state chosen, as long as the initial states possess identical relevant macroscopic quantities.

\subsection{Entanglement entropy dynamics}\label{sec:entropy dynamics}
In this subsection, we aim to investigate the dynamics of entanglement entropy (EE) for various initial states. The EE is a commonly used measure to analyze the ergodic-MBL transition in disorder-driven MBL systems. The bipartite entanglement between a subsystem $ A $ and the rest of the system $ B $ can be effectively quantified by the von Neumann EE
\begin{equation}\label{eq:ee}
	S(t) = -\Tr(\rho_{A}\ln\rho_{A}),
\end{equation}
where $ \rho_{A}=\Tr_{B}(\dyad{\Psi(t)}) $, $\Tr_{B}$  denotes trace with respect
to degrees of freedom of subsystem $B$ and $\ket{\Psi(t)}$ represents the evolved state of the system. We will analyze the half-chain EE where the subsystem $ A $ is chosen to be a block of $ L/2 $ contiguous sites. In ergodic phases, the eigenstates exhibit thermal behavior, resulting in the von Neumann EE being equal to the thermodynamic entropy. As a consequence, the EE scales extensively, following a volume law. In contrast, in conventional MBL phases, the eigenstates can be expressed as product states through quasilocal unitary transformations. This characteristic implies that the von Neuman EE is proportional to the area of the surface between the two subsystems, that is, it obeys the area law scaling.

In Fig.~\ref{fig:entropy}, we show the time evolution of EE $ S(t) $ with several field strengths $ \gamma $ and interacting strengths $ U $. Figs.~\ref{fig:entropy}(a1-a3) and (b1-b3) depict the results for the SDW initial state, while Figs.~\ref{fig:entropy}(c1-c3) and (d1-d3) correspond to the DDW state. In Figs.~\ref{fig:entropy}(a1-a2) and (c1-c2), the $S(t)$ decrease to a stable value for all field gradient $\gamma $. These results reveal that initial states with different energy (the SDW and DDW configurations) can exhibit locally identical dynamics over extended time periods. Especially, in the Mott regime, the EE dynamics $S(t)$ of both SR and LR interacting systems exhibit a noticeable slowdown.

However, we can find that the EE dynamics $S(t)$ for the N{\'e}el initial condition is very different from the results in Fig.~\ref{fig:entropy}. As depicted in Figs.~\ref{fig:imbalance_entropy} (c1-c2) and (d1-d2), when $\gamma \leq \gamma_{c}$, the EE $S(t)$ undergoes a ballistic spreading during the initial evolution and eventually converges to the Page value~\cite{page1993average}, indicating the presence of ergodic dynamics. With increasing the potential strength $\gamma$, the logarithmic growth of EE is observed in Figs.~\ref{fig:imbalance_entropy} (c1-c2) and (d1-d2). This nonergodic behavior is characteristic of Stark MBL ~\cite{pino2014entanglement,luitz2016extended,singh2017effect,yao2020many,lerose2020origin,mohdeb2023entanglement}. Meanwhile, in Fig.~\ref{fig:imbalance_entropy} (c3), we see that the value $S(t)$ of the strongly SR interacting system approaches zero at a long time for $\gamma \geqslant 2 $. For $\gamma =1 $, the entropy exhibits a slowdown behavior with a relatively larger finite value compared to the previous case. This observation suggests the presence of an ergodic-Stark MBL transition in the Mott regime for the strongly interacting SR system. However, as shown in Fig.~\ref{fig:imbalance_entropy} (d3), the entropy displays a similar slowdown behavior with nearly identical finite values for all values of  $\gamma$. These results indicate that the LR interacting system remains in a Stark many-body localized state under strong interaction conditions, regardless of the linear potential strength.

\section{Conclusion}\label{sec:conclusion}
In this study, we theoretically studied the role of interaction strength on the ergodic-Stark MBL transition in one-dimensional spinless fermions systems with a linear potential. We considered both SR and LR interactions in our investigation. By varying the interaction strength, we aimed to understand how it affects the Stark MBL transition in these systems. The main finding of our study is that the interaction strength has a significant impact on disorder-free systems. Specifically, based on the phase diagrams of the mean gap ratio $ \expval{r} $ and the MIPR $\expval{\mathcal{I}}$, we observed that for systems with SR interactions, there exists a qualitative symmetry between the weak and strong interaction limits. However, this symmetry is not present in the case of LR interactions. In the case of LR interactions, the system consistently exhibits Stark MBL phases at strong interaction, irrespective of the strength of the linear potential. This disparity highlights the distinct behavior of SR and LR interactions in disorder-free systems and indacate the robustness of Stark MBL even in the presence of LR interactions. We also study the dynamics of imbalance and entanglement with various initial states using time-dependent variational principle (TDVP) numerical methods. Our analysis uncovered that the dynamical quantities exhibit a significant dependence on the initial conditions. This observation suggests that the fragmentation of the Hilbert-space prevents the system from undergoing thermalization. In summary, our work provides an avenue to investigate MBL in disorder-free systems that involve LR interactions.


\section*{Acknowledgments}
We thank Xiaolong Deng and DinhDuy Vu for fruitful discussions and helpful suggestions. We thank Weilei Zeng and Lyuwen Fu for the technical support on using the Qlab server. The ED and TDVP simulations were performed using the QuSpin~\cite{weinberg2017quspin,weinberg2019quspin} and ITensor libraries\cite{fishman2022itensor}, respectively. This work is supported by National Key Basic Research Program of China (No.~2020YFB0204800), Key Research Projects of Zhejiang Lab (Nos. 2021PB0AC01 and 2021PB0AC02), and the National Science Foundation of China (Grant No.~12204432).

\bibliography{Localization}
\end{document}